\documentclass{article}

\PassOptionsToPackage{square, numbers}{natbib}

\usepackage[preprint]{nips_2018}

\usepackage[utf8]{inputenc} 
\usepackage[T1]{fontenc}    
\usepackage{hyperref}       
\usepackage{url}            
\usepackage{booktabs}       
\usepackage{amsfonts}       
\usepackage{amsmath}
\usepackage{nicefrac}       
\usepackage{microtype}      
\usepackage[pdftex]{graphicx} 

\title{An Insight into the Dynamics and State Space Modelling of a 3-D Quadrotor}

\author{
  Rahul Vigneswaran K\thanks{\url{https://rahulvigneswaran.github.io}}\hspace{1mm} \thanks{Work done while interning at CEN, Amrita Vishwa Vidyapeetham, India.} \\    
  Department of Mechanical Engineering 
  \\Amrita Vishwa Vidyapeetham
  \\Amritapuri, India\\
  \texttt{rahulvigneswaran@gmail.com} 
  \And
  Soman KP \\
  Center for Computational Engineering and Networking (CEN) \\
  Amrita School of Engineering, Amrita Vishwa Vidyapeetham \\
  Coimbatore, India\\
  \texttt{kp\_soman@amrita.edu} \\
}

\begin{document}

\maketitle
\begin{abstract}
  Drones have gained popularity in a wide range of field ranging from aerial photography, aerial mapping, and investigation of electric power lines. Every drone that we know today is carrying out some kind of control algorithm at the low level in order to manoeuvre itself around. For the quadrotor to either control itself autonomously or to develop a high-level user interface for us to control it, we need to understand the basic mathematics behind how it functions. This paper aims to explain the mathematical modelling of the dynamics of a 3 Dimensional quadrotor. As it may seem like a trivial task, it plays a vital role in how we control the drone. Also, additional effort has been taken to explain the transformations of the drone’s frame of reference to the inertial frame of reference.
\end{abstract}

\section{Introduction}

        The popularization of the drone technology has rendered the need for understanding its dynamics and answering the question of \emph{how to control according it to our desirability?} as vital. Drones are formally known as Unmanned Aerial Vehicles (UAVs). It can either be controlled wirelessly by pilots on base stations or can autonomously execute a pre-programmed task. They are very in size and configurations. The drone studied in this paper belongs to the category of micro-drones with 4 symmetrical rigid rotors. Even though the quadrotors have become common, the underlying mathematics is still complicated. This is so, because, quadrotors are \emph{highly coupled}, that is, the fluid dynamics of propellers influence each other. Since its dynamics involve multiple inputs and outputs, it is considered a multivariate problem and call for the requirement of designing a non-linear mathematical model. It must be noted that the system is actually under-actuated. The term under-actuated means, when the number of actuators is less than the degree of freedom. Here too, the drone can move along x,y,z axes and can rotate about the x,y,z axes which makes up for a total of 6 DOF but the quadrotor, as the name suggests is equipped with only 4 actuators, hence effectively making it as an underactuated system. Further on how to overcome this and manoeuvre through all 6 DOF will be discussed later on in this paper. Section 2 gives a deeper understanding of rigid body transformations (both translational and rotational). Section 3 introduces Euler angles from basics and move onto derive a method to find the derivatives of them. Section 4 explains the assumptions involved in the modelling and further proceeds to explaining the kinematics model. Then it shows the design of a dynamics model from scratch by discussing the various forces and aerodynamic moments that affects the model and derives a state-space model for it.

\section{Rigid body Transformations}

Due to the reason that the drone is always in constant motion, it becomes necessary to understand its current position with respect to its previous position or how a set of mutually orthogonal vectors can be mapped to another set of mutually orthogonal vectors. This can be achieved by using \emph{Rigid body transformations} \cite{Brockett1984}. For the purpose of generalization, an arbitrary body is taken into account instead of the quadrotor itself.

\begin{figure}
  \centering
  \includegraphics[width=0.8\linewidth]{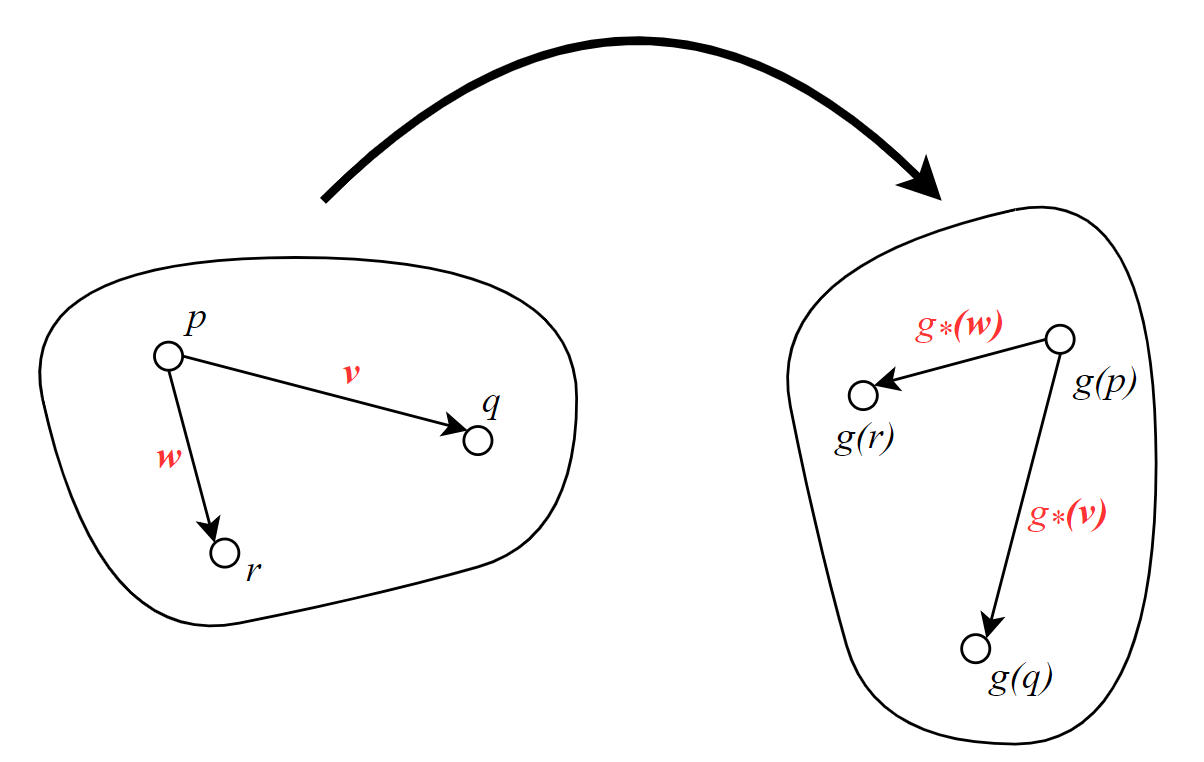}
  \caption{Arbitary points $p,q$ and their corresponding vectors \textbf{\emph{V,W}} are being mapped on a rigid body.}
\end{figure}

For example, let us take the case that is shown in Figure 1. When a point in $p$ is displaced, it is denoted by $g(p)$.\footnote{When a point is transformed, it is denoted by $g$} We must note that displacement is essentially a transformation of points. Similarly, $q$ is displaced to $g(q)$. Now consider the vector $\textbf{\emph{V}}$ from $p$ to $q$. As the points $p$, $q$ moves to $g(p)$, $g(q)$, the $\textbf{\emph{V}}$ is mapped to $g*(\textbf{\emph{V}})$.\footnote{When a vector is transformed, it is denoted by $g_*$} A transformation is valid only when it satisfies the following conditions.

\begin{itemize}
    \item A set of mutually orthogonal vectors after getting mapped must remain mutually orthogonal.
    \item The transformation $g$ must preserve lengths.
    \[\left\| {g(p) - g(q)} \right\| = \left\| {p - q} \right\|\]
    \item The transformation g* must preserve cross products.
    \[{g_*}(\emph{V}) \times {g_*}(\emph{W}) = {g_*}(\emph{V} \times \emph{W})\]
\end{itemize}

\begin{figure}
  \centering
  \includegraphics[width=0.8\linewidth]{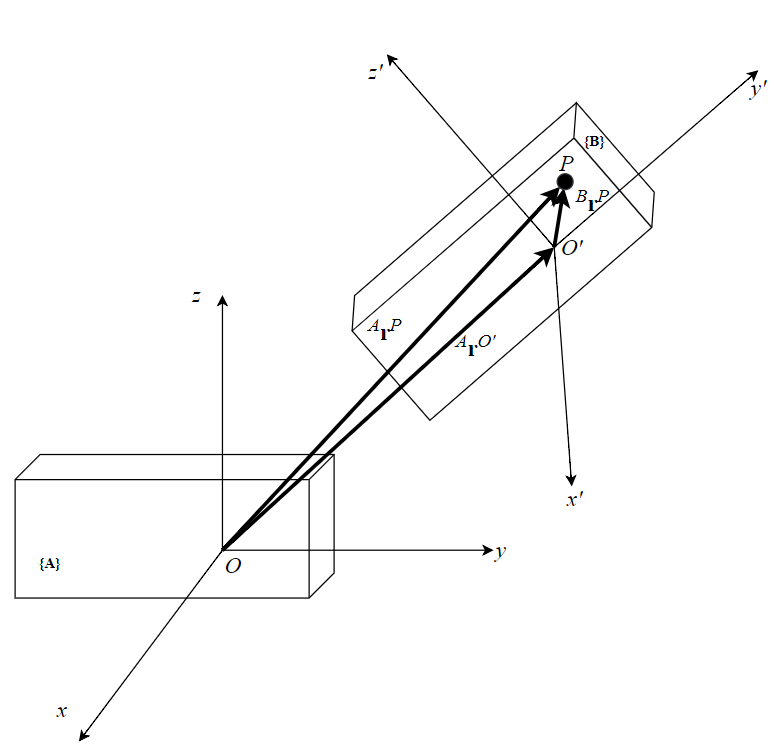}
  \caption{Mapping of two rigid bodies $\{A\}$ and $\{B\}$ with different frame of references.}
\end{figure}

Now let us take two rigid bodies $\{A\}$ and $\{B\}$ in a configuration as shown in Figure 2 and try mapping them a point $P$ on ${B}$ to ${A}$. The body ${A}$ has $x-y-z$ frame of reference with origin $O$ attached to it while the body $\{B\}$ has $x'-y'-z'$ frame of reference with origin $O'$ attached to it. This is a two-stepped process. As we may understand from Figure 2, the mapping involves a rotation mapping and translation mapping. In order to map the $x-y-z$ frame to $x'-y'-z'$ frame, it must be rotated once and then translated. The operations can be carried out the other way too.

\[{}^A{r^P} = {}^A{R_B}{}^B{r^P} + {}^A{r^{O'}}\tag{1}\] 

The equation (1) gives the transformation of point $P$ from frame of reference of $\{B\}$ to $\{A\}$. ${}^A{\textbf{\emph{r}}^P}$ represents the transformation of position vector of $P$ in $\{B\}$ to position vector of $P$ in $\{A\}$ while ${}^B{\textbf{\emph{r}}^P}$ represents the relative position of $\{B\}$ in reference to $\{A\}$. ${}^A{\textbf{\emph{r}}^B}$ represents the rotation of $\{B\}$ in reference to $\{A\}$ and ${}^A{\textbf{\emph{r}}^{O'}}$ represents the translation of origin of $\{B\}$ in reference to $\{A\}$. Therefore, the equation (1) first indicates the rotation transformation of $\{B\}$ in reference to $\{A\}$ and then the translation transformation of the resulting body.

\subsection{Rotational Transformation}

It is essential to understand the basic nature of the rotational transformation to obtain intuitiveness for the rest of the paper. For the sake of simplicity let's take a 2-D configuration as shown in Figure 3 and later extrapolate it to 3-D configuration. From it, the following components of the $V$ and $V'$ can be obtained.

\begin{figure}
  \centering
  \includegraphics[width=0.8\linewidth]{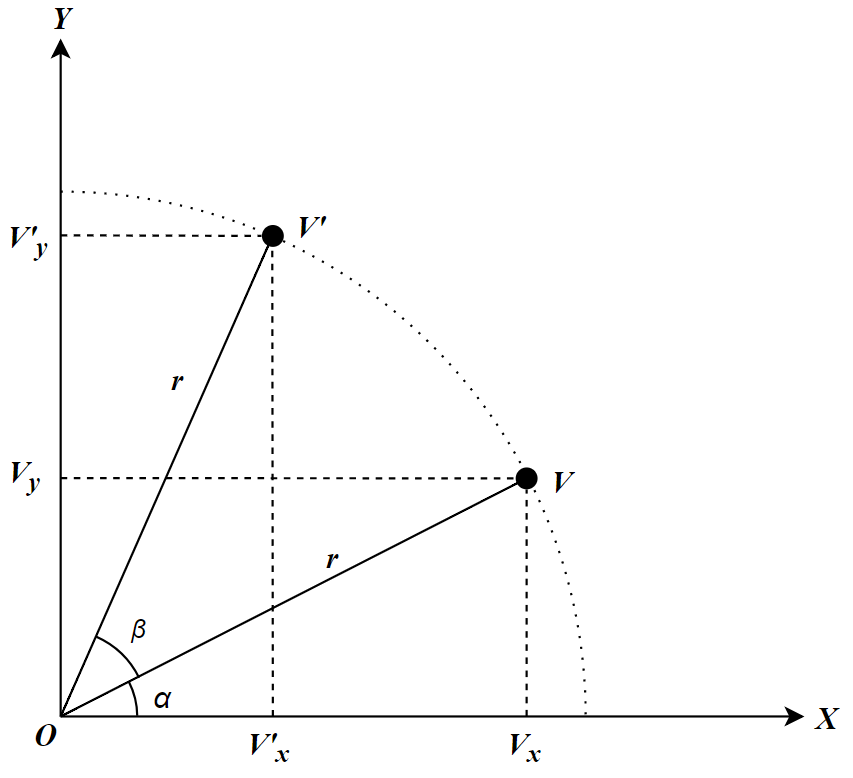}
  \caption{A $x-y$ configuration in which a point from position $V$ is rotated to position $V'$.}
\end{figure}

\[{V_x} = r*\cos \alpha \tag{2}\]
\[{V_y} = r*\sin \alpha \tag{3}\]
\\
\[{{V'}_x} = r*\cos (\alpha  + \beta )\tag{4}\]
\[{{V'}_y} = r*\sin (\alpha  + \beta )\tag{5}\]

Now trigonometric identities are used to break down the equations (4) and (5) to obtain the following,

\[{{V'}_x} = r*\cos \alpha \cos \beta  - r*\sin \alpha \sin \beta \tag{6}\]
\[{{V'}_y} = r*\sin \alpha \cos \beta  - r*\cos \alpha \sin \beta \tag{7}\]

Then by substituting equations (2) and (3) in equations (6) and (7), we get,

\[{{V'}_x} = {V_x}*\cos \beta  - {V_y}*\sin \beta \tag{8}\]
\[{{V'}_y} = {V_y}*\cos \beta  - {V_x}*\sin \beta \tag{9}\]

This when we convert into matrix notation, it gives the rotation of $V$ about an arbitrary point $O$. 

\[\left[ {\begin{array}{*{20}{c}}
  {{{V'}_x}}&{{{V'}_y}} 
\end{array}} \right] = \left[ {\begin{array}{*{20}{c}}
  {{V_x}}&{{V_y}} 
\end{array}} \right]\left[ {\begin{array}{*{20}{c}}
  {\cos \beta }&{\sin \beta } \\ 
  { - \sin \beta }&{\cos \beta } 
\end{array}} \right]\]

Therefore, a 2-D rotation is defined by an angle and an origin. In 3-D, unlike 2-D, it is defined by an angle and a rotational axis. To obtain the rotational transformation in 3-D, as the axis of rotation is restricted to one of the three major axis, on component always remains constant. Then, by following the same procedure as in 2-D, the rotational transformation about x-axis, y-axis and z-axis are obtained.
\\\\
The rotation around x-axis,
\[\left[ {\begin{array}{*{20}{c}}
  {{{V'}_x}}&{{{V'}_y}}&{{{V'}_z}} 
\end{array}} \right] = \left[ {\begin{array}{*{20}{c}}
  {{V_x}}&{{V_y}}&{{V_z}} 
\end{array}} \right]\left[ {\begin{array}{*{20}{c}}
  1&0&0 \\ 
  0&{\cos \beta }&{\sin \beta } \\ 
  0&{ - \sin \beta }&{\cos \beta } 
\end{array}} \right]\]
The rotation around y-axis,
\[\left[ {\begin{array}{*{20}{c}}
  {{{V'}_x}}&{{{V'}_y}}&{{{V'}_z}} 
\end{array}} \right] = \left[ {\begin{array}{*{20}{c}}
  {{V_x}}&{{V_y}}&{{V_z}} 
\end{array}} \right]\left[ {\begin{array}{*{20}{c}}
  {\cos \beta }&0&{ - \sin \beta } \\ 
  0&1&0 \\ 
  {\sin \beta }&0&{\cos \beta } 
\end{array}} \right]\]
The rotation around z-axis,
\[\left[ {\begin{array}{*{20}{c}}
  {{{V'}_x}}&{{{V'}_y}}&{{{V'}_z}} 
\end{array}} \right] = \left[ {\begin{array}{*{20}{c}}
  {{V_x}}&{{V_y}}&{{V_z}} 
\end{array}} \right]\left[ {\begin{array}{*{20}{c}}
  {\cos \beta }&{\sin \beta }&0 \\ 
  { - \sin \beta }&{\cos \beta }&0 \\ 
  0&0&1 
\end{array}} \right]\]

\section{Euler angles}
Leonhard Euler showed that any rotation can be described by three successive rotations \cite{Euler1776} about linearly independent axes and those three angles of rotations are called as roll ($\phi$), pitch ($\theta$) and yaw ($\psi$). Figure 4 illustrates these angles in reference to a glider. The Euler angles $\phi$, $\theta$, $\psi$ can be determined from the rotational transformation matrix. For the purpose of explanation let us take a $Z-Y-X$ rotational transformation. This means that the body is first rotated angle about Z-axis, then angle about Y-axis and then angle about X-axis. Equation (10) is the resulting transformation matrix after $Z-Y-X$ rotation.

\begin{figure}
  \centering
  \includegraphics[width=0.8\linewidth]{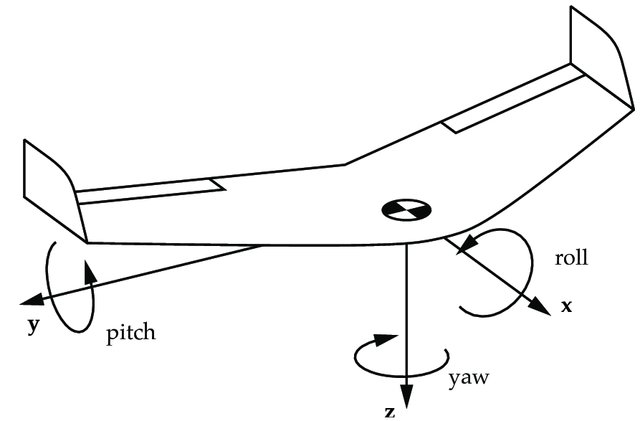}
  \caption{Roll($\phi$), pitch($\theta$), yaw($\psi$) in reference to a glider.}
\end{figure}

\[\begin{gathered}
  R = {R_z}(\psi ){R_y}(\theta ){R_x}(\phi ) \hfill \\
   \quad= \left[ {\begin{array}{*{20}{c}}
  {\cos \psi }&{\sin \psi }&0 \\ 
  { - \sin \psi }&{\cos \psi }&0 \\ 
  0&0&1 
\end{array}} \right]\left[ {\begin{array}{*{20}{c}}
  {\cos \theta }&0&{ - \sin \theta } \\ 
  0&1&0 \\ 
  {\sin \theta }&0&{\cos \theta } 
\end{array}} \right]\left[ {\begin{array}{*{20}{c}}
  1&0&0 \\ 
  0&{\cos \phi }&{\sin \phi } \\ 
  0&{ - \sin \phi }&{\cos \phi } 
\end{array}} \right] \hfill \\
   \quad= \left[ {\begin{array}{*{20}{c}}
  {\cos \theta \cos \psi }&{\cos \psi \sin \theta \sin \phi  - \cos \phi \sin \psi }&{\sin \phi \sin \psi  + \cos \phi \cos \psi \sin \theta } \\ 
  {\cos \theta \sin \psi }&{\cos \phi \cos \psi  + \sin \theta \sin \phi \sin \psi }&{\cos \phi \sin \theta \sin \psi  - \cos \psi \sin \phi } \\ 
  { - \sin \theta }&{\cos \theta \sin \phi }&{\cos \theta \cos \phi } 
\end{array}} \right] \hfill \\
   \quad= \left[ {\begin{array}{*{20}{c}}
  {{R_{11}}}&{{R_{12}}}&{{R_{13}}} \\ 
  {{R_{21}}}&{{R_{22}}}&{{R_{23}}} \\ 
  {{R_{31}}}&{{R_{32}}}&{{R_{33}}} 
\end{array}} \right] \hfill \\ 
\end{gathered}\tag{10}\] 

Every element in the matrix (10) is a combination of two or more angles except for $R_{31}$. By using it as starting equation, we get the following.
\[{R_{31}} =  - \sin \theta \]
\[\theta  =  - {\sin ^{ - 1}}({R_{31}})\tag{11}\]
As we may understand from the equation (11), there are 2 possibilities of $\theta$,

\[{\theta _1} =  - {\sin ^{ - 1}}({R_{31}})\tag{12}\]
\[{\theta _2} = \pi  - {\theta _1} = \pi  + {\sin ^{ - 1}}({R_{31}})\tag{13}\]

From this step on, there are several cases that must be taken into account.\\ \\
\textbf{Case 1.1 :} $( - 1 < \left| {{R_{31}}} \right| < 1)$ and $(\cos \theta  \ne 0)$ 

To find the value of $\phi$, we must observe that,

\[\frac{{{R_{32}}}}{{{R_{33}}}} = \tan \phi \]

Now we use this equation to solve for $\phi$ by using the $\operatorname{atan2}$\footnote{An $\operatorname{atan2(y,x)}$ is an implementation of $\operatorname{atan}(y,x)$ function that solves the sign ambiguity caused, by taking the ratios and signs of $y$ and $x$ into consideration. Also, $\operatorname{atan} (y,x) = \operatorname{atan} \left( {\frac{y}{x}} \right) = {\tan ^{ - 1}}\left( {\frac{y}{x}} \right)$.} function, as

\[\phi  = \operatorname{atan2}({R_{32}},{R_{33}})\tag{14}\]

If $\cos \theta  > 0$, then the equation (14) holds. However, when $\cos \theta  < 0$, then $\phi  = \operatorname{atan2}( - {R_{32}}, - {R_{33}})$. This can be bypassed by using the equation as,

\[\phi  = \operatorname{atan2}\left( {\frac{{{R_{32}}}}{{\cos \theta }},\frac{{{R_{33}}}}{{\cos \theta }}} \right)\tag{15}\]

So, by substituting the equations (12) and (13) in (15), we get,

\[{\phi _1} = \operatorname{atan2}\left( {\frac{{{R_{32}}}}{{\cos {\theta _1}}},\frac{{{R_{33}}}}{{\cos {\theta _1}}}} \right)\tag{16}\]
\[{\phi _2} = \operatorname{atan2}\left( {\frac{{{R_{32}}}}{{\cos {\theta _2}}},\frac{{{R_{33}}}}{{\cos {\theta _2}}}} \right)\tag{17}\]

To find the corresponding values of $\psi$, similar analysis holds good. We observe that,

\[\frac{{{R_{21}}}}{{{R_{11}}}} = \tan \psi \]

We solve for $\psi$ using an equation similar to (14), 

\[\psi  = \operatorname{atan2}\left( {\frac{{{R_{21}}}}{{\cos \theta }},\frac{{{R_{11}}}}{{\cos \theta }}} \right)\tag{18}\]

Now, by substituting the equations (12) and (13) in (18), we get,

\[{\psi _1} = \operatorname{atan2}\left( {\frac{{{R_{21}}}}{{\cos {\theta _1}}},\frac{{{R_{11}}}}{{\cos {\theta _1}}}} \right)\tag{19}\]
\[{\psi _2} = \operatorname{atan2}\left( {\frac{{{R_{21}}}}{{\cos {\theta _2}}},\frac{{{R_{11}}}}{{\cos {\theta _2}}}} \right)\tag{20}\]

Therefore, for the case where $( - 1 < \left| {{R_{31}}} \right| < 1$ and $\cos \theta  \ne 0)$, we get two sets of triplet Euler angles,

\[\begin{gathered}
  ({\phi _1},{\theta _1},{\psi _1}) \hfill \\
  ({\phi _2},{\theta _2},{\psi _2}) \hfill \\ 
\end{gathered} \]

Both of these solutions are absolutely valid.
\\\\
\textbf{Case 1.2 :} $( - 1 < \left| {{R_{31}}} \right| < 1)$ and $(\cos \theta  = 0)$ 

In this case the elements ${R_{11}}$,${R_{21}}$,${R_{32}}$ and ${R_{33}}$ will become zero causing problems\footnote{$\left( {\frac{0}{0}} \right)$ is indeteminate as it can hold any value. Therefore the equation cannot be solved to obtain the values of $\phi$ and $\psi$.} when solving the method used in case 1 and doesn't constrain the values $\phi$,$\psi$. As a result,equations (14) and (18) becomes,

\[\begin{gathered}
  \phi  = \operatorname{atan} 2\left( {\frac{0}{0},\frac{0}{0}} \right) \hfill \\
  \psi  = \operatorname{atan} 2\left( {\frac{0}{0},\frac{0}{0}} \right) \hfill \\ 
\end{gathered} \]

Therefore, we must use different elements of the rotational transformation matrix (10) to find the solution.\\ \\

\textbf{Case 2.1 :} $({R_{31}}=+1)$ 

Since we know that, $({R_{31}} = +1)$. From it, we can get $\theta  = \left( {\frac{\pi }{2}} \right)$. Therefore, the rotational transformation matrix (10) becomes,

\[R = \left[ {\begin{array}{*{20}{c}}
  0&{\cos \psi \sin \phi  - \cos \phi \sin \psi }&{\sin \phi \sin \psi  + \cos \phi \cos \psi } \\ 
  0&{\cos \phi \cos \psi  + \sin \phi \sin \psi }&{\cos \phi \sin \psi  - \cos \psi \sin \phi } \\ 
  { - 1}&0&0 
\end{array}} \right]{\text{ }}\]
As a result, $\psi$ and $\phi$ have an infinite number of valid solutions.

\textbf{Case 2.2 :} $({R_{31}} = -1)$ 

Similar to the case 2.1, since we know that, $({R_{31}} = -1)$. From it, we can get $\theta  =  - \left( {\frac{\pi }{2}} \right)$. Therefore, the rotational transformation matrix (10) becomes,

\[R = \left[ {\begin{array}{*{20}{c}}
  0&{ - \cos \psi \sin \phi  - \cos \phi \sin \psi }&{\sin \phi \sin \psi  - \cos \phi \cos \psi } \\ 
  0&{\cos \phi \cos \psi  - \sin \phi \sin \psi }&{ - \cos \phi \sin \psi  - \cos \psi \sin \phi } \\ 
  1&0&0 
\end{array}} \right]{\text{ }}\]
As a result, $\psi$ and $\phi$ have an infinite number of valid solutions. In both the case 2.1 and 2.2, we have found that $\psi$ and $\phi$ are linked. This phenomenon is called \emph{Gimbal lock}. Although in this case, there is an infinite number of solutions to the problem, in practice, one is often interested in finding one solution. For this task, it is convenient to set $\psi = 0$ and compute $\phi$.

It is interesting to note that there is always more than one sequence of rotations about the three principle axes that result in the same orientation of an object. As an example, consider a book laying on a table face up in front of you. Define the x-axis as to the right, the y-axis as away from you, and the z-axis up. A rotation of $\pi$ radians about the y-axis will turn the book so that the back cover is now facing up. Another way to achieve the same orientation would be to rotate the book $\pi$ radians about the x-axis, and then $\pi$ radians about the z-axis. Thus, there is more than one way to achieve the desired rotation.

\subsection{Derivative of Euler angles}
In order to form the non-linear state space equation later on in this paper, it is necessary that we derive the derivatives of Euler angles. This subsection is dedicated to deriving an equation to obtain derivates of Euler angles ($\dot \phi ,\dot \theta ,\dot \psi $) from angular velocities ($p,q,r$) while undergoing a $Z-Y-X$ rotational transformation. Angular velocities in body frame can be expressed in terms of rate of change of the Euler angles, as follows,

\[\omega  = p\hat i + q\hat j + r\hat k = \pmb{\dot \phi}  + \pmb{\dot \theta}  + \pmb{\dot \psi} \tag{21}\]

The first rotational transformation is ${X_1},{Y_1},{Z_1} \to {X_2},{Y_2},{Z_2}$ along Z-axis. In this ${Z_1}={Z_2}$ which implies that ${\hat k_1} = {\hat k_2}$. Therefore,

\[\pmb{\dot \psi}  = \dot \psi {\hat k_1} = \dot \psi {\hat k_2}\tag{22}\]

The second rotational transformation is ${X_2},{Y_2},{Z_2} \to {X_3},{Y_3},{Z_3}$ along Y-axis. In this ${Y_2}={Y_3}$ which implies that ${\hat j_2} = {\hat j_3}$. Therefore,

\[\pmb{\dot \theta}  = \dot \theta {\hat j_2} = \dot \theta {\hat j_3}\tag{23}\]

The third rotational transformation is ${X_3},{Y_3},{Z_3} \to {X},{Y},{Z}$\footnote{${X},{Y},{Z}$ refers to the inertial/ground frame of reference.} along Z-axis. In this ${X_3}={X}$ which implies that ${\hat i_3} = {\hat i}$. Therefore,

\[\pmb{\dot \phi}  = \dot \phi {\hat i_3} = \dot \phi {\hat i}\tag{24}\]

Now by combining equations (21), (22), (23) and (24), we get,
\[\omega  = p\hat i + q\hat j + r\hat k = \pmb{\dot \phi}  + \pmb{\dot \theta}  + \pmb{\dot \psi}  = \dot \phi \hat i + \dot \theta {\hat j_3} + \dot \psi {\hat k_2}\tag{25}\]

In the transformation ${X_2},{Y_2},{Z_2} \to {X_3},{Y_3},{Z_3}$,
\[\left[ {\begin{array}{*{20}{c}}
  {{U_2}} \\ 
  {{V_2}} \\ 
  {{W_2}} 
\end{array}} \right] = \left[ {\begin{array}{*{20}{c}}
  {\cos \theta }&0&{\sin \theta } \\ 
  0&1&0 \\ 
  { - \sin \theta }&0&{\cos \theta } 
\end{array}} \right]\left[ {\begin{array}{*{20}{c}}
  {{U_3}} \\ 
  {{V_3}} \\ 
  {{W_3}} 
\end{array}} \right]\tag{26}\]

\[\left[ {\begin{array}{*{20}{c}}
  {{{\hat i}_2}} \\ 
  {{{\hat j}_2}} \\ 
  {{{\hat k}_2}} 
\end{array}} \right] = \left[ {\begin{array}{*{20}{c}}
  {\cos \theta }&0&{\sin \theta } \\ 
  0&1&0 \\ 
  { - \sin \theta }&0&{\cos \theta } 
\end{array}} \right]\left[ {\begin{array}{*{20}{c}}
  {{{\hat i}_3}} \\ 
  {{{\hat j}_3}} \\ 
  {{{\hat k}_3}} 
\end{array}} \right]\tag{27}\]\\

In the transformation ${X_3},{Y_3},{Z_3} \to X,Y,Z$,

\[\left[ {\begin{array}{*{20}{c}}
  {{U_3}} \\ 
  {{V_3}} \\ 
  {{W_3}} 
\end{array}} \right] = \left[ {\begin{array}{*{20}{c}}
  1&0&0 \\ 
  0&{\cos \phi }&{ - \sin \phi } \\ 
  0&{\sin \phi }&{\cos \phi } 
\end{array}} \right]\left[ {\begin{array}{*{20}{c}}
  U \\ 
  V \\ 
  W 
\end{array}} \right]\tag{28}\]

\[\left[ {\begin{array}{*{20}{c}}
  {{{\hat i}_3}} \\ 
  {{{\hat j}_3}} \\ 
  {{{\hat k}_3}} 
\end{array}} \right] = \left[ {\begin{array}{*{20}{c}}
  1&0&0 \\ 
  0&{\cos \phi }&{ - \sin \phi } \\ 
  0&{\sin \phi }&{\cos \phi } 
\end{array}} \right]\left[ {\begin{array}{*{20}{c}}
  {\hat i} \\ 
  {\hat j} \\ 
  {\hat k} 
\end{array}} \right]\tag{29}\]
\\
Since, we already know that (26) and (28) are true, if we replace the velocity vector with unit vectors like in (27) and (29), they still hold true. 

From (27) we can find the expression for ${\hat k_2}$ in ${\hat i}$, ${\hat j}$, ${\hat k}$ and it is as follows,

\[{\hat k_2} =  - \sin \theta {\hat i_{^3}} + \cos \theta {\hat k_3}\]

But from (29) we know that, ${\hat k_3} = \sin \phi \hat j + \cos \phi \hat k$ and ${\hat i_3} = \hat i$, so,

\[{\hat k_2} =  - \sin \theta \hat i + \cos \theta \sin \phi \hat j + \cos \theta \cos \phi \hat k\tag{30}\]

Similarly, the expression for ${\hat j_3}$ in ${\hat i}$, ${\hat j}$, ${\hat k}$ can be derived from (29) and it is as follows,
\[{\hat j_3} = \cos \phi \hat j - \sin \phi \hat k\tag{31}\]

Now by substituiting (30) and (31) in (25), we get,
\[p\hat i + q\hat j + r \hat k = \dot \phi \hat i + \dot \theta (\cos \phi \hat j - \sin \phi \hat k) + \dot \psi (- \sin \theta \hat i + \cos \theta \sin \phi \hat j + \cos \theta \cos \phi \hat k)\tag{32}\]

From (32), we can derive the following values,

\[\begin{gathered}
  p = \dot \phi  - (\sin \theta )\dot \psi  \hfill \\
  q = (\cos \phi )\dot \theta  + (\cos \theta \sin \phi )\dot \psi  \hfill \\
  r = (\cos \theta \cos \phi )\dot \psi  - (\sin \phi )\dot \theta  \hfill \\ 
\end{gathered} \]

By rearranging the above in matrix form, we get,

\[\left[ {\begin{array}{*{20}{c}}
  p \\ 
  q \\ 
  r 
\end{array}} \right] = \left[ {\begin{array}{*{20}{c}}
  1&0&{ - \sin \theta } \\ 
  0&{\cos \phi }&{\cos \theta \sin \phi } \\ 
  0&{ - \sin \phi }&{\cos \theta \cos \phi } 
\end{array}} \right]\left[ {\begin{array}{*{20}{c}}
  {\dot \phi } \\ 
  {\dot \theta } \\ 
  {\dot \psi } 
\end{array}} \right]\]

Or,

\[\left[ {\begin{array}{*{20}{c}}
  {\dot \phi } \\ 
  {\dot \theta } \\ 
  {\dot \psi } 
\end{array}} \right] = \left[ {\begin{array}{*{20}{c}}
  1&{\sin \phi \tan \theta }&{\cos \phi \tan \theta } \\ 
  0&{\cos \phi }&{ - \sin \phi } \\ 
  0&{\sin \phi \sec \theta }&{\cos \phi sec\theta } 
\end{array}} \right]\left[ {\begin{array}{*{20}{c}}
  p \\ 
  q \\ 
  r 
\end{array}} \right]\tag{33}\]

The equation (33) can be used to obtain the derivative of Euler angles when we know the angular velocities.

\section{System Modelling of Quadrotor}

Let’s us take the configuration shown in Figure 5 for modelling the system. In this, the propellers 1 and 3 rotate in the same direction while propellers 3 and 4 rotate in opposite direction. This is done so, to overcome the imbalance caused moments of the rotating propellers. A quadrotor is an underactuated system. It has only 4 rotors facing in the same direction in order to manoeuvre through all 6 DOFs. This underactuation is tackled by executing two or more operations (roll, pitch or yaw) at a time to move in the desired direction. Let us take the task of a quadrotor moving forward. When a quadrotor is hovering and is stationary, it has to produce enough overall thrust force (say $F$) to equalize the gravitational force (say $G$) acting on it, that is $F = G$. So, in order to move forward, it has to make a pitch $\theta$. This creates a forward horizontal component $Fsin\theta $ as shown in Figure 6 which enables the quadrotor to move forward. But doing so reduces the vertical component to $Fcos\theta $. As we know from earlier, $Fcos\theta  \ne G$. So, the quadrotor will lose altitude and overcome it, we must increase the thrust force F accordingly to both move forward and not lose altitude. Similarly, all the other operations are executed with a combination of roll, pitch and yaw rotations to overcome the hurdle of underactuation. There are a few assumptions taken into account while modelling the configuration given in Figure 5 to make the modelling easier and they are : 

\begin{itemize}
    \item It is rigid, that is there are no other moving parts other than the rotor. 
    \item It has a symmetrical design.
    \item The propellers are equally spaced from the centre of the body.
    \item Various masses present in quadrotor (battery, circuit boards, etc.) are distributed equally with a centre of mass coinciding with the frame’s centre.
\end{itemize}

\begin{figure}
  \centering
  \includegraphics[width=0.8\linewidth]{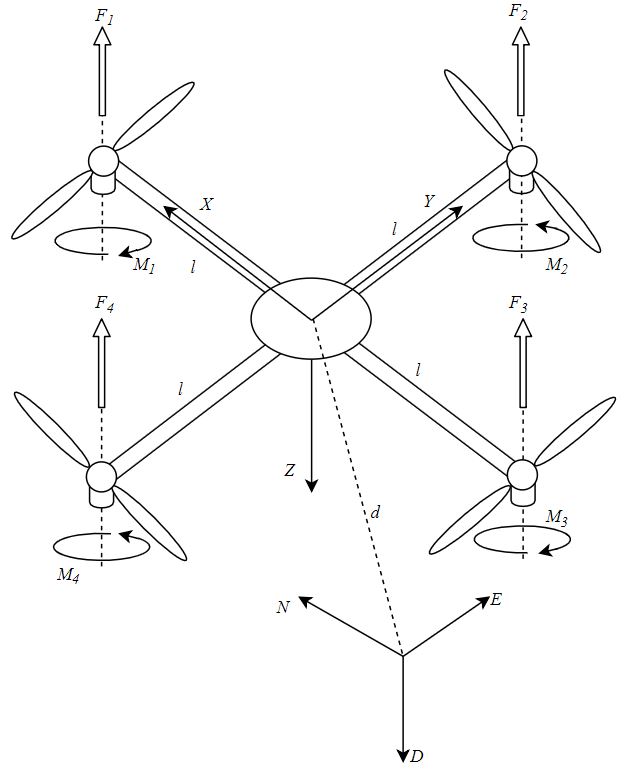}
  \caption{Force diagram of a conventional quadrotor.}
\end{figure}
\begin{figure}
  \centering
  \includegraphics[width=0.8\linewidth]{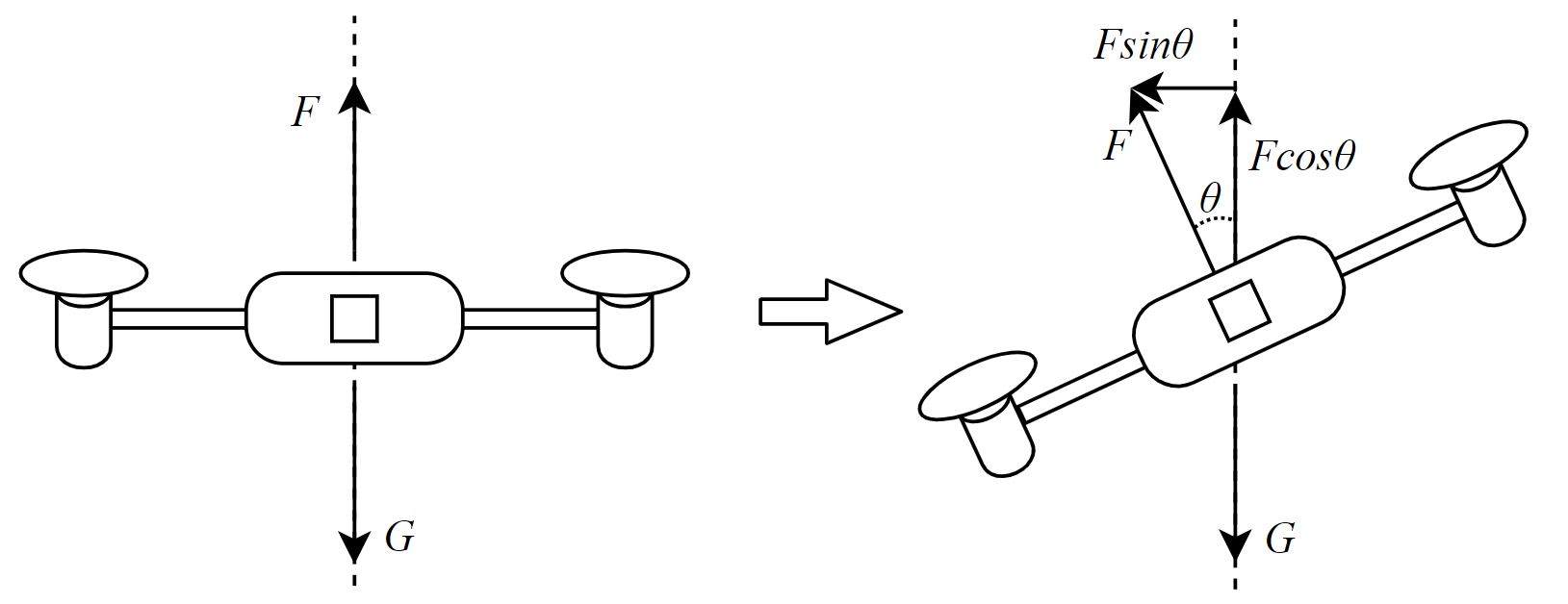}
  \caption{Thrust force diagram of a quadrotor moving in forward direction.}
\end{figure}

\subsection{Kinematic Model}
The first step to model a system is to define the coordinate frames and do the necessary transformations. As shown in Figure 5, the earth inertial frame axes are N, E, D and the body frame axes are X, Y, Z. The inertial frame is fixed to a specific place and the notations N-E-D represent North, East and Downwards respectively. For the body frame of reference, the origin is located at the centre of the quadrotor’s body with X, Y and Z axes pointing towards propellers 1, 2 and ground respectively. The orientation of the quadcopter is defined by the three Euler angles ($\phi, \theta, \psi$).

The thrust force is defined from the body from of reference and we need a transformation matrix (${}^W{R_B}$ - World frame of reference to Body frame of reference) to obtain that. First the rotation transformation is done about Z-axis followed by Y-axis and X-axis. By doing so, we get ,

\[\begin{gathered}
 {}^W{R_B} = {R_z}(\psi ){R_y}(\theta ){R_x}(\phi ) \hfill \\
   \quad= \left[ {\begin{array}{*{20}{c}}
  {\cos \psi }&{\sin \psi }&0 \\ 
  { - \sin \psi }&{\cos \psi }&0 \\ 
  0&0&1 
\end{array}} \right]\left[ {\begin{array}{*{20}{c}}
  {\cos \theta }&0&{ - \sin \theta } \\ 
  0&1&0 \\ 
  {\sin \theta }&0&{\cos \theta } 
\end{array}} \right]\left[ {\begin{array}{*{20}{c}}
  1&0&0 \\ 
  0&{\cos \phi }&{\sin \phi } \\ 
  0&{ - \sin \phi }&{\cos \phi } 
\end{array}} \right] \hfill \\
   \quad= \left[ {\begin{array}{*{20}{c}}
  {\cos \theta \cos \psi }&{\cos \psi \sin \theta \sin \phi  - \cos \phi \sin \psi }&{\sin \phi \sin \psi  + \cos \phi \cos \psi \sin \theta } \\ 
  {\cos \theta \sin \psi }&{\cos \phi \cos \psi  + \sin \theta \sin \phi \sin \psi }&{\cos \phi \sin \theta \sin \psi  - \cos \psi \sin \phi } \\ 
  { - \sin \theta }&{\cos \theta \sin \phi }&{\cos \theta \cos \phi } 
\end{array}} \right] \hfill \\
\end{gathered}\tag{34}\] 

But we need a transformation from the body frame of reference to world frame of reference which can be obtained by inverting ${}^W{R_B}$. By doing so, we get,

\[^B{R_W} = \left[ {\begin{array}{*{20}{c}}
  {\cos \theta \cos \psi }&{\cos \theta \sin \psi }&{ - \sin \theta } \\ 
  {\cos \psi \sin \theta \sin \phi  - \cos \phi \sin \psi }&{\cos \phi \cos \psi  + \sin \theta \sin \phi \sin \psi }&{\cos \theta \sin \phi } \\ 
  {\sin \phi \sin \psi  + \cos \phi \cos \psi \sin \theta }&{\cos \phi \sin \theta \sin \psi  - \cos \psi \sin \phi }&{\cos \theta \cos \phi } 
\end{array}} \right]{\text{ }}\tag{35}\]

\subsection{Dynamic Model}
The quadrotor's motion can be divided into two subsystems - translational motion (X, Y and Z) and rotational motion (roll, pitch and yaw). The forces and moments acting on the quadrotor will be investigated using Newton-Euler formalism. In Figure 5, $F_1$, $F_2$, $F_3$ and $F_4$ are the thrust forces acting on the quadrotor while $M_1$, $M_2$, $M_3$ and $M_4$ are the moments caused by the rotation of the propellers. Also, ${\tau _x}$, ${\tau _y}$ and ${\tau _z}$ are the torques caused by the thrust forces about X, Y and Z axes respectively. 

In some places, for the sake of alignment, 
$\left[ {\begin{array}{*{20}{c}}
  A \\ 
  B \\ 
  C 
\end{array}} \right]$ is written as ${\left[ {\begin{array}{*{20}{c}}
  A&B&C 
\end{array}} \right]^T}$.

\subsubsection{Translational Equations of Motion}
According to Newton's law, 

\[\begin{gathered}
  {{\vec F}_\omega } = \frac{d}{{dt}}(m.\vec v) \\ 
   = m.\dot v \\ 
\end{gathered}\tag{36}\]

where, ${\vec F_\omega } = \vec G - \vec T$. $G$ is the gravity and $T$ is the thrust generated by the propellers. As $\vec T$ is described in body frame of reference, equation (35) is used to transform it to world frame of reference and the expression (36) can be expressed as,

\[\dot v = \frac{1}{m}\left[ {\left[ {\begin{array}{*{20}{c}}
  0 \\ 
  0 \\ 
  {mg} 
\end{array}} \right] - {}^B{R_W}\left[ {\begin{array}{*{20}{c}}
  0 \\ 
  0 \\ 
  T 
\end{array}} \right]} \right]\tag{37}\]

Now substitute (35) in (37), to get,

\[\dot v = \frac{1}{m}\left[ {\begin{array}{*{20}{c}}
  { - T(\cos \phi \sin \theta \cos \psi  + \sin \phi \sin \psi )} \\ 
  { - T(\cos \phi \sin \theta \sin \psi  - \sin \phi \cos \psi )} \\ 
  {mg - T(\cos \phi \cos \theta )} 
\end{array}} \right]\tag{38}\]

This vector equation can be written in terms of its components as,

\[\begin{gathered}
  \ddot x =  - \frac{T}{m}(\cos \phi \sin \theta \cos \psi  + \sin \phi \sin \psi ) \\ 
  \ddot y =  - \frac{T}{m}(\cos \phi \sin \theta \sin \psi  - \sin \phi \cos \psi ) \\ 
  \ddot z = g - \frac{T}{m}(\cos \phi \cos \theta ) \\ 
\end{gathered} \tag{39}\]

\subsubsection{Rotational Equations of Motion}
In the inertial frame, the moment can be defined as the time derivative of the angular momentum and can be expressed as follows,

\[{\vec M_\omega } = \frac{{d\vec L}}{{dt}} = \frac{d}{{dx}}(\vec I \bullet \vec \omega )\tag{40}\]

where, $\vec I$ is the body's inertia tensor.

\[\vec I = \left[ {\begin{array}{*{20}{c}}
  {{I_{xx}}}&{{I_{xy}}}&{{I_{xz}}} \\ 
  {{I_{yx}}}&{{I_{yy}}}&{{I_{yz}}} \\ 
  {{I_{zx}}}&{{I_{zy}}}&{{I_{zz}}} 
\end{array}} \right] = \left[ {\begin{array}{*{20}{c}}
  {{I_{xx}}}&0&0 \\ 
  0&{{I_{yy}}}&0 \\ 
  0&0&{{I_{zz}}} 
\end{array}} \right]]\tag{41}\]

In equation (41), the off-diagonal terms are zero due to the symmetric mass distribution assumption as mentioned in Section 4. 

According to Euler's rotational equation,

\[{\vec M_b} = \left( {\vec I \bullet \dot \vec \omega } \right) + \vec \omega  \times \left( {\vec I \bullet \vec \omega } \right) = \left[ {\begin{array}{*{20}{c}}
  {{\tau _x}} \\ 
  {{\tau _y}} \\ 
  {{\tau _z}} 
\end{array}} \right]\tag{42}\]

where, $\dot \vec \omega  = \left[ {\begin{array}{*{20}{c}}
  {\dot p} \\ 
  {\dot q} \\ 
  {\dot r} 
\end{array}} \right]$ and the right hand side is a vector of applied torques. 

Also, by simplifying $\vec \omega  \times \left( {\vec I \bullet \vec \omega } \right)$, we get,

\[\vec \omega  \times \left( {\vec I \bullet \vec \omega } \right) = \left[ {\begin{array}{*{20}{c}}
  p \\ 
  q \\ 
  r 
\end{array}} \right] \times \left[ {\begin{array}{*{20}{c}}
  {{I_{xx}}p} \\ 
  {{I_{yy}}q} \\ 
  {{I_{zz}}r} 
\end{array}} \right] = \left[ {\begin{array}{*{20}{c}}
  {({I_{zz}} - {I_{yy}})qr} \\ 
  {({I_{yy}} - {I_{zz}})rp} \\ 
  {({I_{yy}} - {I_{xx}})pq} 
\end{array}} \right]\tag{43}\]

Now, by substituting (43) in (42), we get,

\[\begin{gathered}
  {\tau _x} = {I_{xx}}\dot p + ({I_{zz}} - {I_{yy}})qr \\ 
  {\tau _y} = {I_{yy}}\dot q + ({I_{yy}} - {I_{zz}})rp \\ 
  {\tau _z} = {I_{zz}}\dot r + ({I_{yy}} - {I_{xx}})pq \\ 
\end{gathered} \tag{44}\]

 Here$p, q, r$ are actually equal to $\dot \phi, \dot \theta, \dot \psi$ since by definition, angular velocity of the body are same as time derivatives of Euler angles of it .When the eqution (44) is rearranged, we get,

\[\begin{gathered}
  \dot p = \ddot \phi = ({\tau _x} + {I_{yy}}qr - {I_{zz}}qr)/{I_{xx}} \\ 
  \dot q = \ddot \theta = ({\tau _y} - {I_{xx}}pr + {I_{xx}}pr)/{I_{yy}} \\ 
  \dot r = \ddot \psi = ({\tau _z} + {I_{xx}}pq - {I_{yy}}pq)/{I_{zz}} \\ 
\end{gathered} \tag{45}\]

Using equation (33), we can get the derivatives of the Euler angles as follows,

\[\begin{gathered}
  \dot \phi  = p + {\text{ }}(r\cos (\phi )\sin (\theta ))/\cos (\theta ){\text{ }} + {\text{ }}(q\sin (\theta )\sin (\phi ))/\cos (\theta ) \hfill \\
  \dot \theta  = q\cos (\phi ) - r\sin (\phi ) \hfill \\
  \dot \psi  = {\text{ }}(r\cos (\phi ))/\cos (\theta ){\text{ }} + {\text{ }}(q\sin (\phi ))/\cos (\theta ) \hfill \\ 
\end{gathered} \tag{46}\]

\subsubsection{Aerodynamic Moment and thrust Force}
As the propeller rotates, it generates a thrust force. This thrust force is directly proportional \cite{Pounds2004} to the square of the rotor speed and can be expressed as,

\[{F_i} =  - {K_a}\omega _i^2\tag{47}\]

Apart from thrust force, the rotating propellers generate aerodynamic moment which is also directly proportional to the square of the rotor speed and can be expressed,

\[{M_i} =  - {K_m}\omega _i^2\tag{48}\]

For the case of low altitude flight, the values of $K_a$ and $K_m$ in expression (47) and (48) can be considered as constants. The directions of the thrust force and aerodynamic moments generated by the propellers is illustrated in the Figure 5. Now, we can take the moment caused by the thrust force generated into account. Let's say that the length of the arms reaching out to the propellers from the centre of the body is $l$. Now, the total moment caused by thrust force about the X-axis is,

\[\begin{gathered}
  {\tau _x} =  - {F_2}l + {F_4}l \hfill \\
   =  - ({K_a}\omega _2^2)d + ({K_a}\omega _4^2)l \hfill \\
   = {K_a}l(\omega _4^2 - \omega _2^2) \hfill \\ 
\end{gathered} \tag{49}\]

Total moment caused by the generated thrust about Y-axis is,

\[\begin{gathered}
  {\tau _y} = {F_1}l - {F_3}l \\ 
   = ({K_a}\omega _1^2)l - ({K_a}\omega _3^2)l \\ 
   = {K_a}l(\omega _1^2 - \omega _3^2) \\ 
\end{gathered} \tag{50}\]

The total moment about Z-axis is influenced only by the aerodynamic moment caused by the propellers (48) but not affected by the thrust force and it can be expressed as,

\[\begin{gathered}
  {\tau _z} = {M_1} - {M_2} + {M_3} - {M_4} \\ 
   = {\text{ }}({K_m}\omega _1^2) - ({K_m}\omega _2^2){\text{ }} + ({K_m}\omega _3^2) - ({K_m}\omega _4^2) \\ 
   = {K_m}(\omega _1^2 - \omega _2^2 + \omega _3^2 - \omega _4^2) \\ 
\end{gathered} \tag{51}\]

The total thrust force can be expressed as,

\[T =  - {K_a}(\omega _1^2 + \omega _2^2 + \omega _3^2 + \omega _4^2)\tag{52}\]

The negative sign in expression (52) indicates that the total thrust force is acting in the negative direction of Z-axis. Also, in this quadrotor configuration, thrust force exists only in the direction of Z-axis.

\subsection{State Space Equation}
In general, the evolution of a dynamical system's states are governed by set of ordinary differential equations. In control problems, these ordinary differential equations are often rearranged into state-space form,

\[\dot X = f(X,U)\tag{53}\]

where, $X$ is a matrix of states and $U$ is a matrix of inputs. The formation of state space equation involves 3 steps :
\begin{itemize}
    \item Identify the governing equations and and find the order ($n$) of it.
    \item Define the state and input vector.
    \item Define the system of first-order differential equations.\\
\end{itemize}

\subsubsection{Governing Equations}

Equations (39) and (45) are the governing equations and are as follows,

\[\begin{gathered}
  \ddot x =  - \frac{T}{m}(\cos \phi \sin \theta \cos \psi  + \sin \phi \sin \psi ) \\ 
  \ddot y =  - \frac{T}{m}(\cos \phi \sin \theta \sin \psi  - \sin \phi \cos \psi ) \\ 
  \ddot z = g - \frac{T}{m}(\cos \phi \cos \theta ) \\ 
\end{gathered} \tag{39}\]

\[\begin{gathered}
  \dot p = \ddot \phi = ({\tau _x} + {I_{yy}}qr - {I_{zz}}qr)/{I_{xx}} \\ 
  \dot q = \ddot \theta = ({\tau _y} - {I_{xx}}pr + {I_{xx}}pr)/{I_{yy}} \\ 
  \dot r = \ddot \psi = ({\tau _z} + {I_{xx}}pq - {I_{yy}}pq)/{I_{zz}} \\ 
\end{gathered} \tag{45}\]

The order ($n$) of the all the governing equation is 2.

\subsubsection{State and Input Vector}

The order of each governing equation is 2. Therefore, each governing equation needs $(n-1)$ states, that is 2 states each. Therefore, in order to define the motion of a quadrotor, practically, we will need $(2\times6)$ states, that is 12 states - $x, y, z, \dot x, \dot y, \dot z, \phi, \theta, \psi, \dot \phi, \dot \theta, \dot \psi$. Therefore the matrix of the states of the system is,

\[\begin{gathered}
  X = {\left[ {\begin{array}{*{20}{c}}
  {{x_1}}&{{x_2}}&{{x_3}}&{{x_4}}&{{x_5}}&{{x_6}}&{{x_7}}&{{x_8}}&{{x_9}}&{{x_{10}}}&{{x_{11}}}&{{x_{12}}} 
\end{array}} \right]^T} \hfill \\
   = {\left[ {\begin{array}{*{20}{c}}
  x&y&z&\phi &\theta &\psi &{\dot x}&{\dot y}&{\dot z}&{\dot \phi }&{\dot \theta }&{\dot \psi } 
\end{array}} \right]^T} \hfill \\ 
\end{gathered} \tag{54}\]

The matrix of the inputs to the system is,

\[\begin{gathered}
  U = \left[ {\begin{array}{*{20}{c}}
  {{U_1}}&{{U_2}}&{{U_3}}&{{U_4}} 
\end{array}} \right] \hfill \\
   = \left[ {\begin{array}{*{20}{c}}
  {\omega _1^2}&{\omega _2^2}&{\omega _3^2}&{\omega _4^2} 
\end{array}} \right] \hfill \\ 
\end{gathered} \tag{55}\]

\subsubsection{System of first-order equations}

The first 6 equations simply related states to each other by finding their derivatives like,

\[\begin{gathered}
  \frac{{d{x_1}}}{{dt}} = \frac{{dx}}{{dt}} = \dot x = {x_7} \\ 
  \frac{{d{x_2}}}{{dt}} = \frac{{dy}}{{dt}} = \dot y = {x_8} \\ 
  \frac{{d{x_3}}}{{dt}} = \frac{{dz}}{{dt}} = \dot z = {x_9} \\ 
  \frac{{d{x_4}}}{{dt}} = \frac{{d\phi }}{{dt}} = \dot \phi  = {x_{10}} \\ 
  \frac{{d{x_5}}}{{dt}} = \frac{{d\theta }}{{dt}} = \dot \theta  = {x_{11}} \\ 
  \frac{{d{x_6}}}{{dt}} = \frac{{d\psi }}{{dt}} = \dot \psi  = {x_{12}} \\ 
\end{gathered} \]

The next 6 equations can be derived by differentiating and rearranging the above 6 first-order differential equations like,

\[\begin{gathered}
  \frac{{d{x_7}}}{{dt}} = \frac{{d\dot x}}{{dt}} = \ddot x \\ 
  \frac{{d{x_8}}}{{dt}} = \frac{{d\dot y}}{{dt}} = \ddot y \\ 
  \frac{{d{x_9}}}{{dt}} = \frac{{d\dot z}}{{dt}} = \ddot z \\ 
  \frac{{d{x_{10}}}}{{dt}} = \frac{{d\dot \phi }}{{dt}} = \ddot \phi  \\ 
  \frac{{d{x_{11}}}}{{dt}} = \frac{{d\dot \theta }}{{dt}} = \ddot \theta  \\ 
  \frac{{d{x_{12}}}}{{dt}} = \frac{{d\dot \psi }}{{dt}} = \ddot \psi  \\ 
\end{gathered} \]

Now, by substituting the values of values for $\ddot x,\ddot y,\ddot z,\dot \phi ,\dot \theta ,\dot \psi ,\ddot \phi ,\ddot \theta ,\ddot \psi $, from the the equations (39), (45) and (46) onto the above 12 first-order differential equationa=s, we get our state space equation in the form $\{\dot X = f(X,U)\}$ as,

\[\left[ {\begin{array}{*{20}{c}}
  {{\dot X_1}} \\ 
  {{\dot X_2}} \\ 
  {{\dot X_3}} \\ 
  {{\dot X_4}} \\ 
  {{\dot X_5}} \\ 
  {{\dot X_6}} \\ 
  {{\dot X_7}} \\ 
  {{\dot X_8}} \\ 
  {{\dot X_9}} \\ 
  {{\dot X_{10}}} \\ 
  {{\dot X_{11}}} \\ 
  {{\dot X_{12}}} 
\end{array}} \right] = \left[ {\begin{array}{*{20}{c}}
  {{f_1}} \\ 
  {{f_2}} \\ 
  {{f_3}} \\ 
  {{f_4}} \\ 
  {{f_5}} \\ 
  {{f_6}} \\ 
  {{f_7}} \\ 
  {{f_8}} \\ 
  {{f_9}} \\ 
  {{f_{10}}} \\ 
  {{f_{11}}} \\ 
  {{f_{12}}} 
\end{array}} \right] = \left[ {\begin{array}{*{20}{c}}
  {\dot x} \\ 
  {\dot y} \\ 
  {\dot z} \\ 
  {\dot \phi } \\ 
  {\dot \theta } \\ 
  {\dot \psi } \\ 
  {\ddot x} \\ 
  {\ddot y} \\ 
  {\ddot z} \\ 
  {\ddot \phi } \\ 
  {\ddot \theta } \\ 
  {\ddot \psi } 
\end{array}} \right] = \left[ {\begin{array}{*{20}{c}}
  {{x_7}} \\ 
  {{x_8}} \\ 
  {{x_9}} \\ 
  {p + {\text{ }}(r\cos (\phi )\sin (\theta ))/\cos (\theta ){\text{ }} + {\text{ }}(q\sin (\theta )\sin (\phi ))/\cos (\theta )} \\ 
  {q\cos (\phi ) - r\sin (\phi )} \\ 
  {(r\cos (\phi ))/\cos (\theta ){\text{ }} + {\text{ }}(q\sin (\phi ))/\cos (\theta )} \\ 
  { - \frac{T}{m}(\cos \phi \sin \theta \cos \psi  + \sin \phi \sin \psi )} \\ 
  { - \frac{T}{m}(\cos \phi \sin \theta \sin \psi  - \sin \phi \cos \psi )} \\ 
  {g - \frac{T}{m}(\cos \phi \cos \theta )} \\ 
  {({\tau _x} + {I_{yy}}qr - {I_{zz}}qr)/{I_{xx}}} \\ 
  {({\tau _y} - {I_{xx}}pr + {I_{xx}}pr)/{I_{yy}}} \\ 
  {({\tau _z} + {I_{xx}}pq - {I_{yy}}pq)/{I_{zz}}} 
\end{array}} \right]\]

Therefore, the non-linear state-space equation for a 3-D quadrotor has been derived. It is non-linear due to the presence $sine$ and $cosine$ terms in it.

\section{Conclusion}
Dynamics of a 3-D quadrotor and its state-space equation has been derived from the grass root level with detailed explanation and the logical reasoning behind every step have been stated clearly. As a future work, the dynamical system that is modelled in this paper can be used in conjunction with concepts of control system theory to make filters and controllers that can make the quadrotor follow a trajectory or tackle any kind of task with precision and robustness using MATLAB. Apart from that, the concepts of Quaternions \cite{Shepperd1978} can be used in the modelling of the system to make the computations easier.

\subsubsection*{Acknowledgments}

This paper includes a collective knowledge majorly from the lectures of \href{https://www.kumarrobotics.org/dr-vijay-kumar/}{Dr.Vijay Kumar} and \href{https://www.sarahtang.net}{Sarah Tang} from University of Pennsylvania and the master thesis work \cite{Lyu2017} of Haifeng Lyu from University of Rhode Island. Also, this paper would have not been possible without the continued guidance of \href{https://www.amrita.edu/faculty/kp-soman}{Dr. KP. Soman} of CEN, Amrita Vishwa Vidyapeetham, India.

\small

\bibliography{nips_2018}
\bibliographystyle{unsrt}

\end{document}